# Network cloning using DNA barcodes

Sergey A. Shuvaev, Batuhan Başerdem, Anthony Zador, Alexei A. Koulakov

*Cold Spring Harbor Laboratory, Cold Spring Harbor, NY 11724*

The ability to measure or manipulate network connectivity is the main challenge in the field of connectomics. Recently, a set of approaches has been developed that takes advantage of next generation DNA sequencing to scan connections between neurons into a set of DNA barcodes. Individual DNA sequences called markers represent single neurons, while pairs of markers, called barcodes contain information about connections. Here we propose a strategy for 'copying' or 'cloning' connectivity contained in barcodes into a clean slate *tabula rasa* network. We show that a one marker one cell (OMOC) rule, which forces all markers with the same sequence to condense into the same neuron, leads to fast and reliable formation of desired connectivity in a new network. We show that OMOC rule yields convergence in a number of steps given by a power law function of the network size. We thus propose that copying network connectivity from one network to another is theoretically possible.

The connections between neurons determine the computations performed by a neural network. In both biological and artificial neural networks, these connections are established and tuned by experience and learning. The learned connections can thus be considered a "summary" of the statistical structure of the experience—data—on which the network was trained. This summary, which can be represented as a list of connections, may be considerably more compact and efficient than the original data. For example, typical deep networks for implementing object detection require tens of millions of connections, derived from data sets containing hundreds of billions pixels, which results in more than 1000-fold compression [1, 2]. It would therefore be more efficient to copy these connections onto a new network than to retrain a new network from scratch.

Most current implementations of artificial neural networks are on digital computers and GPUs [2]. On these architectures, connections are stored explicitly and therefore straightforward to extract and copy into a new network. However, in biological networks, there is no central repository for connections, so reading out the connections of a network and copying them into a new network represents a difficult challenge.

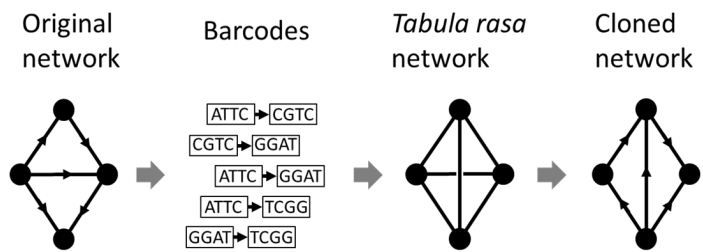

**Figure 1.** Network cloning as a way to copy connectivity from one network to another. The original network is read out into a set of barcodes carrying information about connections. Each half of the barcode (marker) represents one of the cells that are connected, while the link represents the direction of the connections. These barcodes are then introduced into a *tabula rasa* network that has no structure. Barcodes are capable to shape the *tabula rasa* network to match the target connectivity.

We have recently proposed a new way to read out neuronal connections using DNA barcodes [3, 4]. In this strategy, individual neurons produce distinguishable pseudo-random DNA identifiers called markers. Pairs of markers, called here barcodes, represent individual synaptic connections. Barcodes are read out using high-throughput sequencing technology, either *in situ* [5] or *ex vivo* after individual neurons are disassociated. This strategy allows to convert connections between neurons into an ensemble of DNA barcodes that can be identified using sequencing methods.

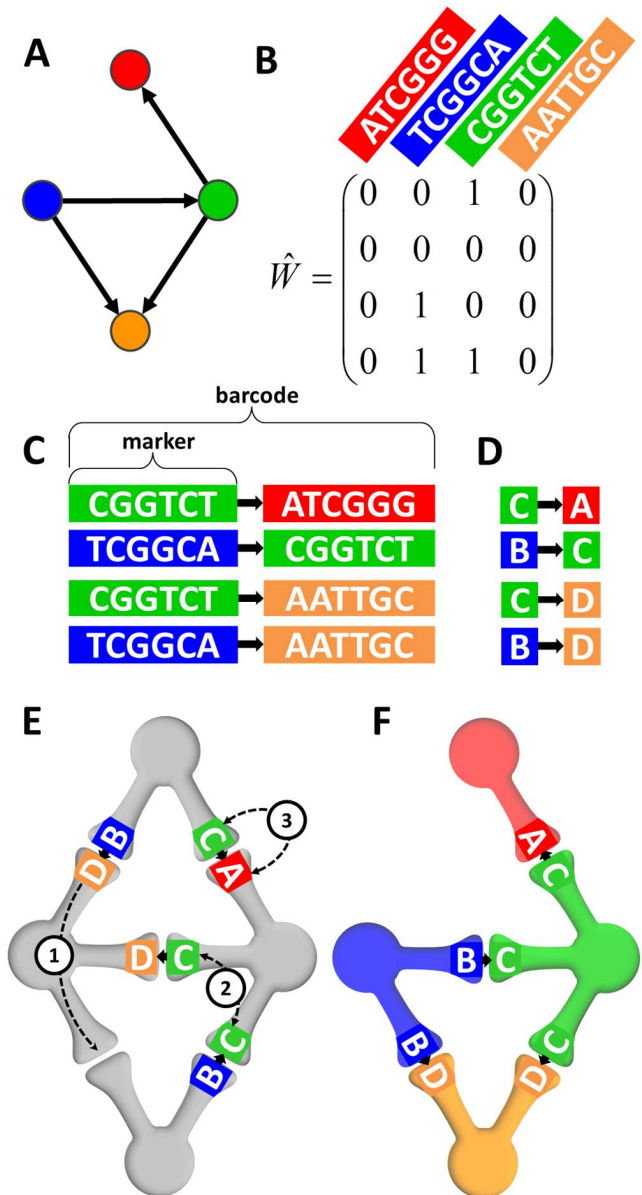

**Figure 2.** One marker one cell (OMOC) rule yields target connectivity. (A) A target network of four cells. (B) The connection (weight) matrix corresponding to the target connectivity. Each network node is associated with a random nucleotide sequence that we call a marker. (C) Barcodes are pairs of markers connected by a directional spacer. Each barcode represents a non-zero entry in the connection matrix in (B). (D) Instead of using full sequences, we will represent markers by unique letters of an alphabet. (E) The barcodes are initially arranged randomly in neurons connected in all-to-all function. The barcodes can move through the network by jumping from synapse to synapse using three moves as illustrated: jump (1), swap (2), and flip (3). (F) The proposed model forces all markers facing every neuron to be the same. This arrangement is called OMOC. Once OMOC solution is achieved, we eliminate all synapses that contain no barcodes, such as the synapse between cells "A" and "B". Our hypothesis is that OMOC solution yields the copying of the original connection matrix.

Here we formulate a different question: Given an ensemble of connections represented by barcodes, can we copy them into a new network? In other words, can original network be cloned? We explore a computational model that simulates the behavior of barcodes introduced into a *tabula rasa* network with unstructured connectivity and test its ability to recreate target connectivity in such networks (Fig. 1). We require that the underlying mechanisms be purely local, i.e. the behavior of each cell and barcode is based on the information available in this cell or in its synapses only. Below we present the model that allows robust copying of connectivity based on such local interaction rules only.

Our model attempts to recreate the target connectivity between neurons (Fig. 2A). The connectivity can be represented as a connection matrix (Fig. 2B). We assume that every network node (neuron) is identified by a unique barcode sequence that we call a marker (Fig. 2B). Marker is a sequence of nucleotides long enough to uniquely label every neuron in the network. The network connectivity is encoded by pairs of markers that we call barcodes. Each barcode contains the marker corresponding to the presynaptic cell, a postsynaptic marker, and a spacer between two markers indicating the connection's direction (Fig. 2C). Thus, each barcode encodes a unique connection between cells. The number of barcodes is equal to the number of non-zero entries in the connection matrix or to the total number of connections in the network. To simplify notations, we represent each marker by an individual letter of an alphabet that is long enough to describe every neuron in the network (Fig. 2D).

The barcodes are introduced into synapses of a *tabula rasa* network that is, initially, all-to-all. Since connectivity in our model is directional (Fig. 2), we assume that, between every two cells, synapses are formed initially in both directions. The full connectivity assumption is made here to simplify the description of network dynamics, however, a more sophisticated synaptic dynamics is also possible [6]. The assumption of all-to-all initial connectivity is also not so far from reality, as cortical neurons typically form a network that is 10% sparse (pairs of neurons that can be physically connected are connected with >10% probability) [7], while, during development, the synapses are overproduced by a factor of 10-20 [6, 8], rendering both potential and developmental connectivity that is close to full [7]. The number of neurons in the *tabula rasa* network is assumed to be equal to the number of nodes in the desired network, i.e. equal to the number of markers.

The barcodes are initially introduced into synapses randomly (Fig. 2E). The barcodes are then allowed to be rearranged in the network via three types of moves. First, we allowed the barcodes to hop from one synapse to another synapse of the same cell (jumps). Second, we allowed two barcodes located in the same neuron to trade their places (swaps). Although a swap can be viewed as a superposition of two jumps, we use swaps as an independent type of moves to avoid issues with the network lingering in metastable states for too long. Finally, each barcode is permitted to be reinserted in the synapse between the same pair of cells in different orientation (flips). In implementing these moves we keep track of the direction of barcodes and synapses, i.e. barcodes are introduced into synapses of the correct orientation. The arrangement of a barcode in the direction opposite to the direction of a synapse is considered to be energetically prohibitive. The moves that we use are also local which means that the barcodes can only be relocated between synapses of the same neuron.

Using this set of moves, we rearrange barcodes in the network implementing one marker – one cell (OMOC) rule. OMOC rule forces all markers (half-barcodes) in the synapses of the same cell, facing this cell, to be the same (Fig. 2F). Thus, in Fig. 2F, *all* markers in the rightmost cell are described by letter C (A, B, C, D is a short-hand notation for much longer nucleotide sequences). Similarly, all markers in the leftmost cell are labeled by letter B. We surmised that if the logic of the interaction of cells and barcodes implements OMOC rule, cells will discover their identity as encoded by markers (half-barcodes). Because every cell in the *tabula rasa* network has a potential to become any cell as defined by the markers, a specific arrangement of barcodes respecting OMOC rule is associated with a symmetry breaking, whereby the network selects one possible assignment of markers into cells out of $N!$ combinations ($N$ is the number of cells in the network equal to the number of markers). We also surmised that if we then eliminate all synapses that are *not* occupied by a barcode, the remaining synapses will implement the target connectivity.

To practically implement OMOC solution, we defined a cost-function, $H$, that is minimized by the barcode dynamics. Although many choices are possible for the cost function, we chose this particular form:

$$H = -(1+\varepsilon)\sum_{c=1}^{N}\sum_{m=1}^{M}(n_{cm})^{\gamma} + \varepsilon\sum_{c=1}^{N}(\sum_{m=1}^{M}n_{cm})^{\gamma} \quad (1)$$

This cost function depends on the distribution of barcodes via matrix $n_{cm}$ defining, for each cell number $c$, the number of markers of type $m$ located on this cell. The cost-function depends on two parameters, $\gamma$ and $\varepsilon$. To implement OMOC rule, the former parameter has to be larger than one (in our simulations, we used $\gamma = 2$). With $\gamma = 2$, the first term of the cost-function can be viewed as the measure of sparseness of marker distribution $n_{cm}$ [9, 10]; minimizing the measure of marker sparseness will yield a single marker dominating each cell, i.e. OMOC solution. The second term in the cost-function defines the penalty for placing a non-dominant marker in each cell. This penalty is controlled by an independent parameter $\varepsilon = 10$ in our simulations. For $\gamma = 2$, the cost function can be written as $H = \sum_{c=1}^{N}\vec{n}_c^T \hat{U} \vec{n}_c$, where $\vec{n}_c$ is the vector of marker abundances in cell number $c$, and $\hat{U} = -(1+\varepsilon)\hat{I} + \varepsilon\hat{Y}$. Here $\hat{Y}$ is the matrix of all ones. Because the diagonal part of matrix $\hat{U}$ ($\sim \hat{I}$) is negative, it facilitates single marker type per cell, while the off-diagonal part ($\sim \hat{Y}$) penaltizes multiple marker types in a cell. Both of these components help achieve OMOC solution.

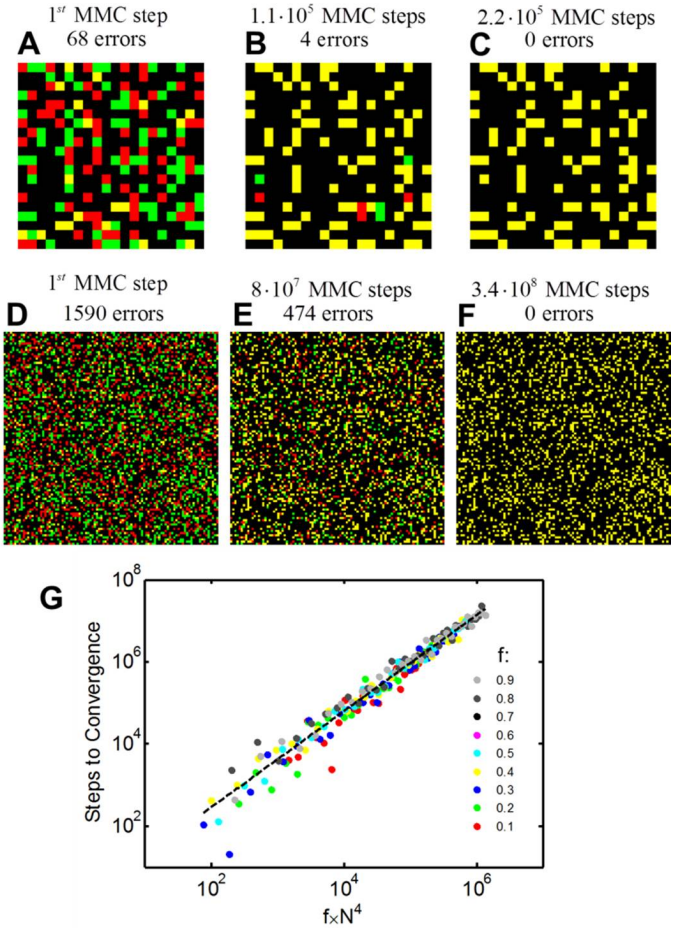

**Figure 3.** OMOC rule allows copying desired network connectivity matrix. Results of a single MMC simulated annealing run for 20x20 (A-C) and 100x100 (D-F) networks. Red/green channels show target/actual connection matrices. Yellow matrices at the end of the simulation run (C and F) indicate a perfect copy. (G) OMOC rule yields target connectivity in a number of steps given by a power law of network size. Number of steps required for convergence as a function of a combination of network parameters ($N$, the size of the network, and $f$, the fraction of non-zero connections). Dashed line is the best linear fit corresponding to the number of steps $N_{steps} \propto (fN^4)^{1.17}$.

The approach based on minimizing a cost-function is one of the ways to quantitatively describe biological processes and has been used successfully to describe establishing connectivity, especially when competition or interdependence between cells is important [11-13]. To minimize the cost function we use a conventional Metropolis Monte Carlo (MMC) simulated annealing procedure [11, 12] and three types of barcode moves as described above. After the cost function is minimized, at the end of MMC procedure, we remove synapses that carry no barcodes. This step is based on the assumption that synapses carrying barcodes are more stable than their barcodeless counterparts. Overall, we hypothesized that, when OMOC solution is reached, the final connectivity between cells will reproduce the connectivity between barcodes.

Our results show that, indeed, OMOC rule yields desired connectivity after several MMC steps (Figure 3). We tested the convergence on a set of randomly generated asymmetric sparse networks, with the fraction of non-zero connections determined by parameter $f$. We find that, even for substantially large connection matrices (Figure 3D-F), the target connectivity can be reached in relatively small number of steps in 100% of cases. To quantify the speed of convergence, for each MMC simulation, we computed the number of attempts to move barcodes before a perfect OMOC solution was achieved, $N_{steps}$. We find that this parameter is well approximated by a power-law function

$$N_{steps} \propto (fN^4)^\alpha \qquad (2)$$

Here $N$ is the number of cells in the network (the size of the connection matrix), $\alpha \simeq 1.17$ is the scaling exponent. This means that OMOC rule yields target connectivity in $\sim N^{4.68}$ steps, i.e. a power law function of the network size. This result suggests that even for large networks, the original connectivity can be reached in a finite, i.e. non-exponential, number of steps. Overall our results show that copying the target connectivity to the new network is possible with OMOC rule and the convergence of this rule is relatively fast.

If one assumes that the scaling exponent $\alpha \simeq 1.17$ is approximately equal to one, the number of steps to convergence can be represented as $N_{steps} \propto BN^2$, where $B = fN^2$ is the total number of barcodes (synapses) in the network. The number of steps that each barcode has to make is given by $N_{steps}/B \propto N^2$. Thus, to a first approximation, each barcode has to explore $\sim N^2$ potential positions before OMOC solution is achieved, the number that is independent on other barcodes. This may explain why the rate of convergence is given by a power law function and is not exponential in the number of neurons.

Here we addressed the question whether connectivity can be copied from one neural network to another. We assumed that the connections are represented by an ensemble of DNA barcodes [3, 4]. We analyzed the dynamics of barcodes introduced into a clean slate *tabula rasa* network. The particular form of dynamics that we considered is described by one marker one cell rule (OMOC), which favors positioning of a single type of DNA sequence in a single neuron. We showed that OMOC dynamics leads to fast and reliable recreation of desired connectivity in the new network. The formation of new connectivity is achieved in a number of steps given by a power law of the network size. Thus, copying connectivity from one neural network to another using DNA barcodes is theoretically possible.

**Methods**

To minimize the cost function (1), we used simulated annealing procedure [11, 12]. We started from a random distribution of barcodes in synapses of a fully connected directed network. Barcodes were relocated between synapses as described, according to Monte Carlo statistical rules. The temperature was gradually lowered from $10^{-2}$ to $10^{-6}$ of the initial value. The number of steps in the algorithm was chosen to be 10 times the value given by equation (2). The probabilities of three operations, jumps, swaps, and flips, were $1-f$, $f-1/N$, and $1/N$ respectively. During each of the first two operations, a barcode was inserted in a random orientation. To compare to the target connectivity, we used a greedy procedure that finds dominant markers for each cell.